\documentclass[11pt]{article}
\pdfoutput=1
\usepackage{jheppubmod}
\usepackage[utf8]{inputenc}
\usepackage{amsmath,amsfonts,amstext,amssymb,mathrsfs,mathtools}
\usepackage{graphicx}
\usepackage{subcaption}
\usepackage{comment}
\usepackage{dutchcal}
\usepackage{color}
\usepackage{cancel}
\usepackage{multicol}
\usepackage{multirow}
\usepackage{graphicx}
\usepackage[sorting=none]{biblatex}
\newcommand{\secref}[1]{Section \ref{#1}}

\newcommand{\figref}[1]{Fig. \ref{#1}}

\newcommand{\ed}{\text{d}}

\newcommand{\p}{\partial}

\begin{document}

\title{$2 \rightarrow 2N$ scattering: Eikonalisation and the Page curve}
\author{Nava Gaddam and}
\author{Nico Groenenboom}
\emailAdd{gaddam@uu.nl}
\emailAdd{n.groenenboom@uu.nl}
\affiliation{Institute for Theoretical Physics and Center for Extreme Matter and Emergent Phenomena, Utrecht University, 3508 TD Utrecht, The Netherlands.}
\date{\today}

\abstract{In the spirit of studying the information paradox as a scattering problem, we pose and answer the following questions: i) What is the scattering amplitude for $N$ particles to emerge from a large black hole when two energetic particles are thrown into it? ii) How long would we have to wait to recover the information sent in? The answer to the second question has long been expected to be Page time, a quantity associated with the lifetime of the black hole. We answer the first by evaluating an infinite number of `ladder of ladders' Feynman diagrams to all orders in $M_{Pl}/M_{BH}$. Such processes can generically be calculated in effective field theory in the black hole eikonal phase where scattering energies satisfy $E M_{BH} \gg M^{2}_{Pl}$. Importantly, interactions are mediated by a fluctuating metric; a fixed geometry is insufficient to capture these effects. We find that the characteristic time spent by the particles in the scattering region (the so-called Eisenbud-Wigner time delay) is indeed Page time, confirming the long-standing expectation. This implies that the entropy of radiation continues to increase, after the particles are thrown in, until after Page time, when information begins to re-emerge.}

\maketitle

\section{Introduction and conclusions}

Shortly after Hawking claimed that black holes led to information loss \cite{Hawking:1974sw, Hawking:1976ra}, Page suggested \cite{Page:1979tc} that in emitting a fraction of their energy, black holes would create $\mathcal{O}\left(G M^{2}\right)$ particles in $\Delta t \sim G^{2} M^{3}$ (Page time). The consequence of this emission would be, provided momentum conservation holds, that the recoil of the black hole would be larger than the size of the black hole itself. Thereby invalidating Hawking's assumption that the black hole remains in its original rest frame as in the semiclassical approximation of a single fixed classical geometry. In addition to other neglected correlations, the position of the recoil could potentially be in precise correlation with the emitted radiation. Later, Page went on to propose that an interesting quantity to track information retrieval from the black hole is the entanglement entropy of the radiation over time \cite{Page:1993wv}. Given a putative quantum gravity theory with a unitary scattering matrix, the entanglement entropy of the radiation must then begin to reduce, in contrast to the thermodynamic entropy, after Page time; this is the Page curve. Inspired by ideas from string theory and gauge/gravity duality, it has been recently suggested that the downturn of the Page curve is already visible in effective field theory owing to additional saddles in the gravitational path integral \cite{Penington:2019npb, Almheiri:2019hni}. An important question now concerns the precise dynamics that govern information retrieval in effective field theory.

\begin{figure}[h!]
\centering
\includegraphics[scale=0.5]{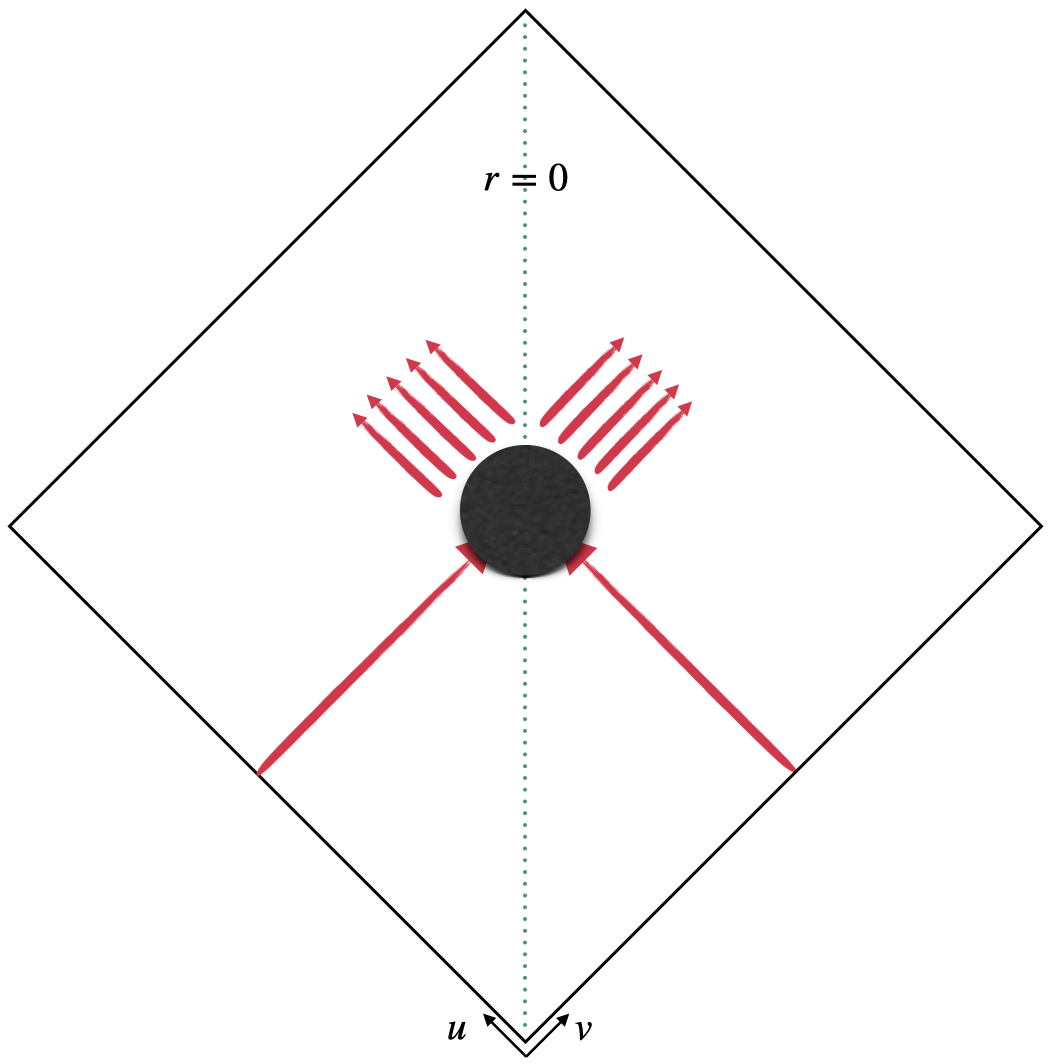}
\caption{A schematic diagram of the two (green) infalling particles into the black hole and the 2N (red) outgoing particles escaping to the future.}
\label{fig:2-2N_schematic}
\end{figure}

\begin{figure}[h!]
\centering
\includegraphics[scale=0.4]{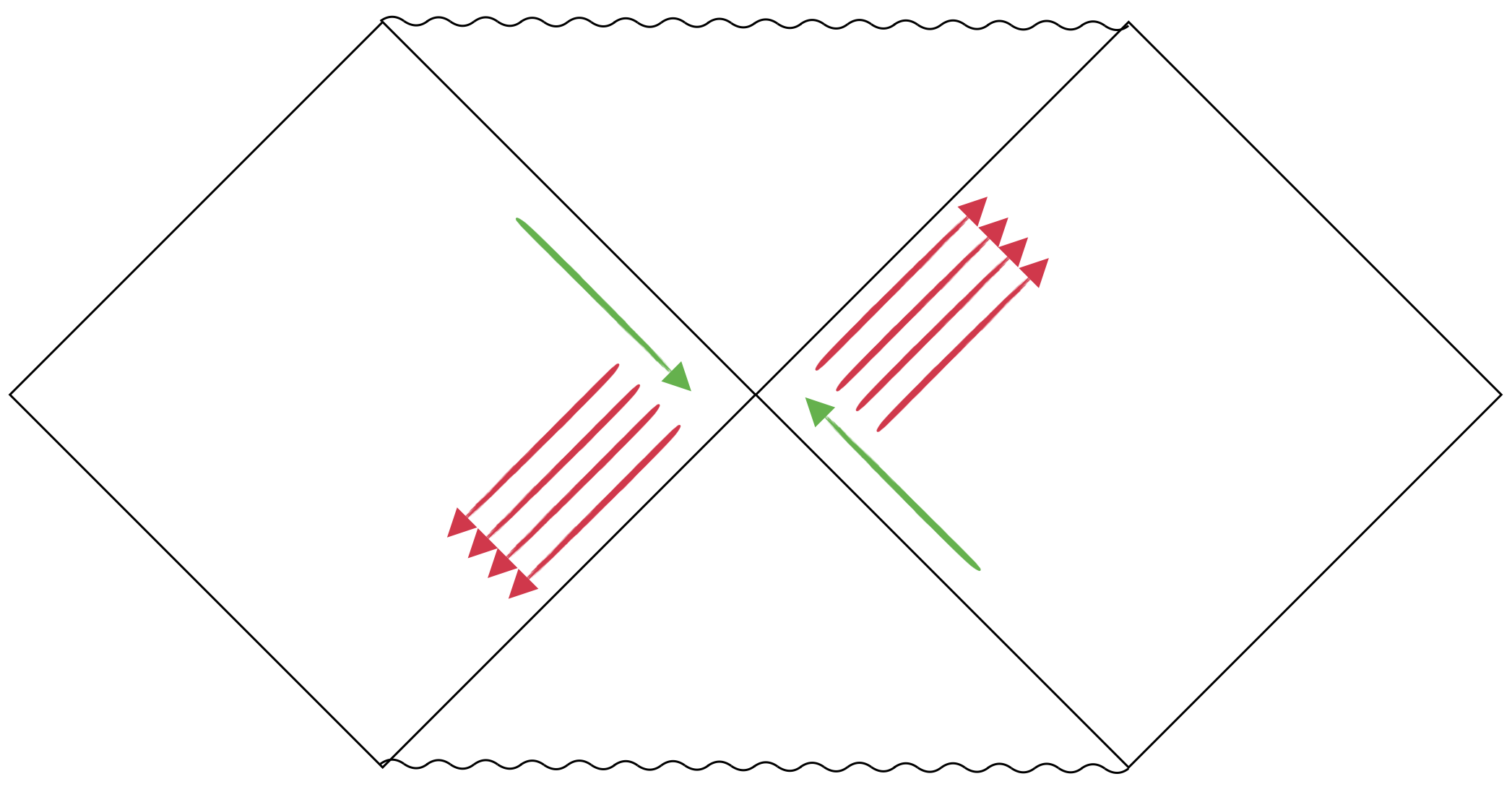}
\caption{The scattering processes of interest can also be represented on the Carter-Penrose diagram of the two-sided eternal black hole: the green and red arrows represent ingoing and outgoing particles respectively. The antipodal identification of \cite{Hooft:2016itl, Betzios:2017krj} on this diagram leads us to the schematic picture in \figref{fig:2-2N_schematic}.}
\label{fig:2-2N_eternalBH}
\end{figure}

In this article, to address this problem, we consider the throwing of two particles into a large semiclassical black hole which subsequently emits an arbitrary number of outgoing particles that escape to future infinity. Such a process is shown in \figref{fig:2-2N_schematic} and \figref{fig:2-2N_eternalBH}. 

\newpage

We now pose the following questions: i) can we compute S-matrix elements for this process within effective field theory? and ii) can we estimate how long particles spend in the scattering region before escaping? We will show in this article that the answer to the first question is affirmative and find an explicit formula for the amplitude. Given such a scattering matrix $S\left(E^{2}\right)$, Eisenbud-Wigner's time delay \cite{Eisenbud:1948, Wigner:1955zz, Martin:1981dea, Bianchi:2012} estimates the time the scattering particles spend in the scattering region before escaping to infinity. If an intermediate metastable bound-state is formed, this time delay can be seen to estimate the lifetime of the said intermediate state. It is given by
\begin{equation}\label{eqn:wigner_timedelay}
\Delta T ~ = ~ \text{Re} \left[- i S^{\dagger} \dfrac{\partial S}{\partial E}\right] \, .
\end{equation}
Using this expression, we will be able to answer the second question too in the affirmative. The answer turns out to depend on the centre of mass energy of the scattering process. Studying tree level $2\rightarrow 2N$ amplitudes, we first find that the amplitudes are very sharply peaked around a maximum value $N_{max}$. The time spent in the scattering region for elastic $2\rightarrow 2$ is the scrambling time. Whereas when particle production is considered ($N>1$), we find the time spent in the scattering region to be approximately $G^{2} M^{3}$, namely Page time. This implies that the entanglement entropy of the radiation continues to grow until Page time for as long as the particles are trapped in the scattering region. After Page time, however, the scattering matrix elements dictate that information about the particles that were thrown in begins to emerge, forcing the entanglement entropy to take a down turn. The dynamics is governed by the interactions between the propagating particles and gravitons arising from metric perturbations. A semiclassical approximation with a single fixed classical geometry misses this physics entirely. These interactions are reliably calculable in a regime of phase space where the centre of mass energies of the scattering particles satisfies 
\begin{equation}\label{eqn:bheikonal}
E \, M_{BH} \gg M^{2}_{Pl} \, .
\end{equation}
The importance of this phase in quantum gravity was anticipated by Veneziano \cite{Veneziano:2012yj}. For large semiclassical black holes, this condition is easily satisfied even with low scattering energies. The scattering matrix approach of 't Hooft's can be seen to establish the importance of this phase insofar as elastic $2\rightarrow 2$ scattering \cite{tHooft:1996rdg, Hooft:2015jea, Hooft:2016itl, Betzios:2016yaq, Betzios:2020wcv} is concerned. Its generalisation to a second quantised formulation \cite{Gaddam:2020rxb, Gaddam:2020mwe, Betzios:2020xuj} allows for the study of scattering processes with particle production that we undertake in this article. 

\paragraph{Organisation of this paper:}  In \secref{sec:eikonalisation}, we first review the Feynman rules governing the scattering processes, then compute tree level $2\rightarrow 2N$ amplitudes to estimate the dominant scattering energies, and further calculate an eikonal resummation of these amplitudes to all orders in $M_{Pl}/M_{BH}$. In \secref{sec:timedelay}, we compute the time spent by the particles in the scattering region before concluding in \secref{sec:outlook}. 

\section{$2\rightarrow 2N$ scattering: the `ladder of ladders' diagrams}\label{sec:eikonalisation}
\begin{figure}[h!]
\begin{subfigure}[h]{0.45\linewidth}
\centering
   \includegraphics[scale=0.38]{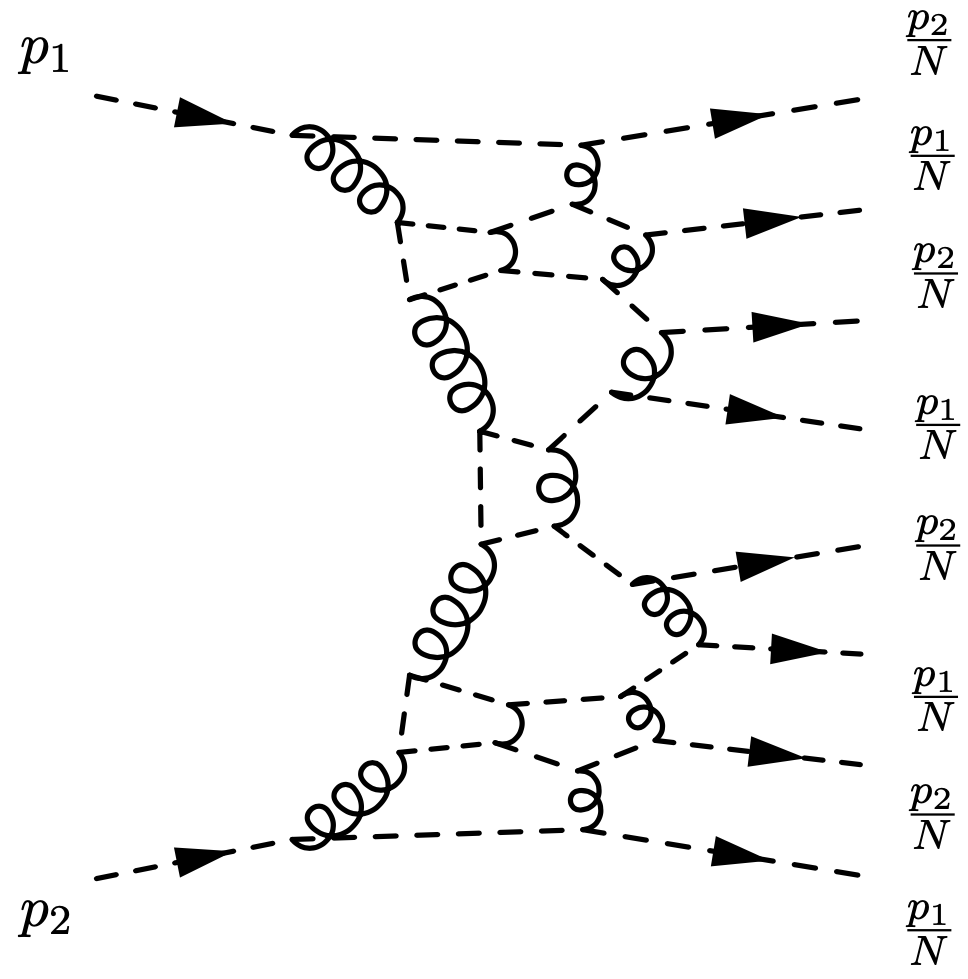}
\caption{The `cobweb' diagram}
\label{fig:2-2N_cobweb}
\end{subfigure}
\hfill
\begin{subfigure}[h]{0.45\linewidth}
\centering
   \includegraphics[scale=0.38]{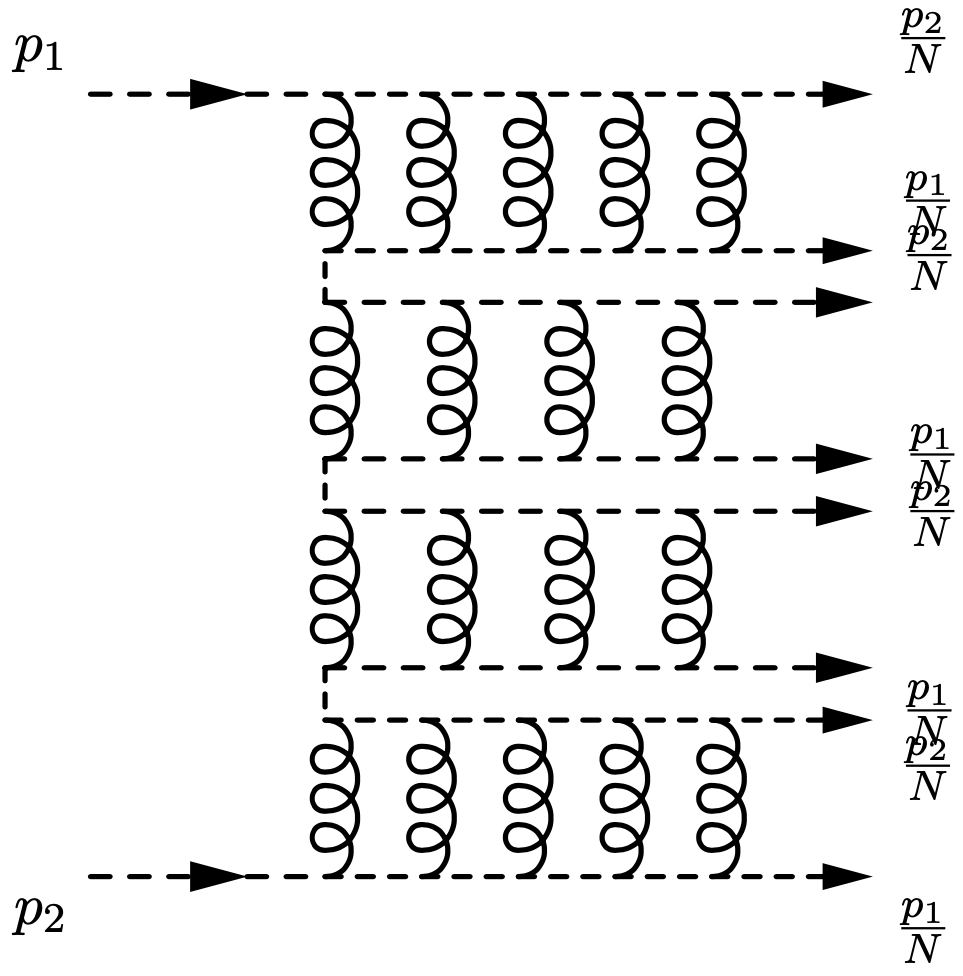}
    \caption{The `ladder of ladders' diagram}
    \label{fig:2-2N_ladders}
\end{subfigure}
\caption{The general $2\rightarrow 2N$ eikonal loop graph resembles a `cobweb' and is difficult to compute. Whereas the calculable graph is what we call the `ladder of ladders' diagram. In both graphs, momentum conservation demands that the initial incoming momentum is distributed among all final outgoing legs. Assuming no special directions, the dominant contribution comes from democratically distributed momentum among all external legs. The Feynman rules  and choice of incoming momenta dictate that the outgoing external legs alternate in momenta, leaving two possible orders. Remarkably, we find that all `ladder of ladder' exchanges can be resummed.}
\end{figure}

The scattering processes of interest are governed by interactions between gravitons on the fluctuating background spacetime and the external particles \cite{Gaddam:2020rxb, Gaddam:2020mwe}. Just as in the elastic $2 \rightarrow 2$ scattering in the black hole eikonal, the first interaction term is the three vertex that couples the graviton to the stress tensor of the minimally coupled scalar field. We will leave the study of higher order interactions for future work. This has the consequence that the scattering processes available to us involve only an even number of in or out going legs. In the language of $2\rightarrow 2$ scattering, this is akin to studying the leading eikonal. However, unlike in the $2\rightarrow 2$ case, where a general eikonal loop diagram is a ladder diagram, a general $2\rightarrow 2N$ diagram is a `cobweb' as shown in \figref{fig:2-2N_cobweb}. Such diagrams are difficult to compute in general. However, a closely related diagram that is easier to calculate is a `ladder of ladders' as shown in \figref{fig:2-2N_ladders}.

\subsection{Feynman rules for the $\ell=0$ modes}
It was shown in \cite{Martel:2005ir, Gaddam:2020rxb, Gaddam:2020mwe} that the even and odd parity graviton modes decouple in the Regge-Wheeler gauge \cite{Regge:1957td}. The quadratic action for the even parity graviton is given by
\begin{align}
S ~ &= ~ \dfrac{1}{4} \sum_{\ell , m} \int\ed^2 x \biggr(\mathfrak{h}^{ab} \Delta^{-1}_{abcd} \mathfrak{h}^{cd} + \mathfrak{h}^{ab} \Delta^{-1}_{ab} \mathcal{K} + \mathcal{K} \Delta^{-1}_{ab} \mathfrak{h}^{ab} + \mathcal{K} \Delta^{-1} \mathcal{K}\biggr) \, ,
\end{align}
where partial wave indices for the fields $\mathfrak{h}_{ab}$ and $\mathcal{K}$ have been suppressed, and the background operators were written in \cite{Gaddam:2020rxb, Gaddam:2020mwe}. As was pointed out in those references and later explained in detail in \cite{Kallosh:2021ors, Kallosh:2021uxa}, there is residual gauge redundancy in these fields for the $\ell=0,1$ modes. In this article, we will focus on the $\ell=0$ mode which, in the Regge-Wheeler gauge takes the form\footnote{In fact, our field $\mathfrak{h}_{ab}$ includes a field redefinition compared to the familiar fields in the Regge-Wheeler gauge $\mathfrak{h}_{ab} = \frac{1}{r A\left(r\right)} h_{ab}$ \cite{Regge:1957td}. This field redefinition is performed in addition to a Weyl rescaling of the light cone metric $g_{ab} = A\left(r\right) \eta_{ab}$ which allows for a definition of kinematic/Mandelstam variables in the near horizon region \cite{Gaddam:2020rxb, Gaddam:2020mwe}.}
\begin{equation}
\mathfrak{h}_{\mu\nu} ~ = ~ \dfrac{r}{A\left(r\right)}\begin{pmatrix}
H_{xx} & H_{xy} & 0 & 0 \\
H_{xyt} & H_{yy} & 0 & 0 \\
0 & 0 & r^{2} A\left(r\right) \mathcal{K} & 0 \\
0 & 0 & 0 & r^{2} A\left(r\right) \mathcal{K} \sin^{2}\left(\theta\right) 
\end{pmatrix} \, .
\end{equation}
Since the angular diffeomorphisms do not contribute to this spherically symmetric mode, it has two gauge redundancies remaining. One may be fixed by choosing $\mathcal{K}=0$. In \cite{Kallosh:2021ors}, it was proposed that the final gauge degree of freedom may be fixed by contracting the graviton with a pair of orthogonal vectors and setting the result to zero. Instead, in the present article, we will choose a different gauge condition that is more convenient, namely tracelessness $\eta^{ab}\mathfrak{h}_{ab}=0$. Since $\eta^{ab}$ is off-diagonal in the light-cone coordinates we are working in, this implies that $H_{xy}=0$. Given that the odd parity mode vanishes identically for the $\ell=0$, the gauge fixed action is entirely given by the even parity mode and can be written as
\begin{align}
S ~ &= ~ \dfrac{1}{4} \int\ed^2 x \mathfrak{h}^{ab} \Delta^{-1}_{abcd} \mathfrak{h}^{cd} \, ,
\end{align}
where the operator $\Delta^{-1}_{abcd}$ is given by
\begin{align}
\Delta^{-1}_{abcd} ~ &= ~ \dfrac{1}{2 R^2} \left(\eta_{ac} x_{[b} \p_{d]} + \eta_{bd} x_{[a} \p_{c]}\right) + \dfrac{1}{2 R^2} \bigr(\eta_{ab}\eta_{cd}-\eta_{ac}\eta_{bd}-\eta_{ac}\eta_{bd}) \, ,
\end{align}
where $R$ is the Schwarzschild radius associated with the background. In comparison to the operator written in \cite{Gaddam:2020rxb, Gaddam:2020mwe}, this one takes the additional tracelessness gauge condition into account for the $\ell=0$ monopole mode. The propagator can be worked out following the steps in \cite{Gaddam:2020rxb, Gaddam:2020mwe} to find
\begin{align}\label{eqn:monopoleProp}
    2 i \mathcal{P}_{abcd} ~ = ~ i R^2 \bigr(\eta_{ab}\eta_{cd}-\eta_{ac}\eta_{bd}-\eta_{ac}\eta_{bd}) \, .
\end{align}
We see that the momentum dependence drops out entirely, which is consistent with the expectation that this mode does not propagate; it can never appear as an external leg in any amplitude. Nevertheless, it is physical in that it contributes to amplitudes in the form of virtual legs. 

For simplicity, we will consider the scattering of massless scalar particles; their effective two dimensional propagators were shown to be \cite{Gaddam:2020rxb, Gaddam:2020mwe}: 
\begin{equation}
\dfrac{-i}{p^{2} + \frac{\ell^{2} + \ell + 1}{R^{2}}} ~ = ~ \dfrac{-i}{p^{2} + \frac{1}{R^{2}}} \quad \text{when $\ell = 0$} \, .
\end{equation}
Although the curvature of the horizon renders the scalar particles with a small effective two-dimensional mass, it was argued in \cite{Gaddam:2020rxb, Gaddam:2020mwe} that the contribution of this effective mass is suppressed in scattering amplitudes. Therefore, we will treat the external particles to be on-shell even as they are chosen to be light-like. The interactions between the virtual gravitons and scalar legs are mediated by the three vertex $i \gamma p_{\mu} p_{\nu}$ governed by the coupling constant $\gamma = \kappa R^{-1}$ where $\kappa = \sqrt{8 \pi G}$.

These rules focus on the dynamics of the near-horizon region. It would be interesting to incorporate effects of the classical potential barrier further away, as was done for $2\rightarrow 2$ scattering in \cite{Betzios:2020xuj}. However, these effects are not important for the $\ell = 0$ modes that we concern ourselves with in this article.

\subsection{Tree level diagrams and the saddle point}\label{sec:tree_saddle}
\begin{figure}[h!]
\centering
    \includegraphics[scale=0.35]{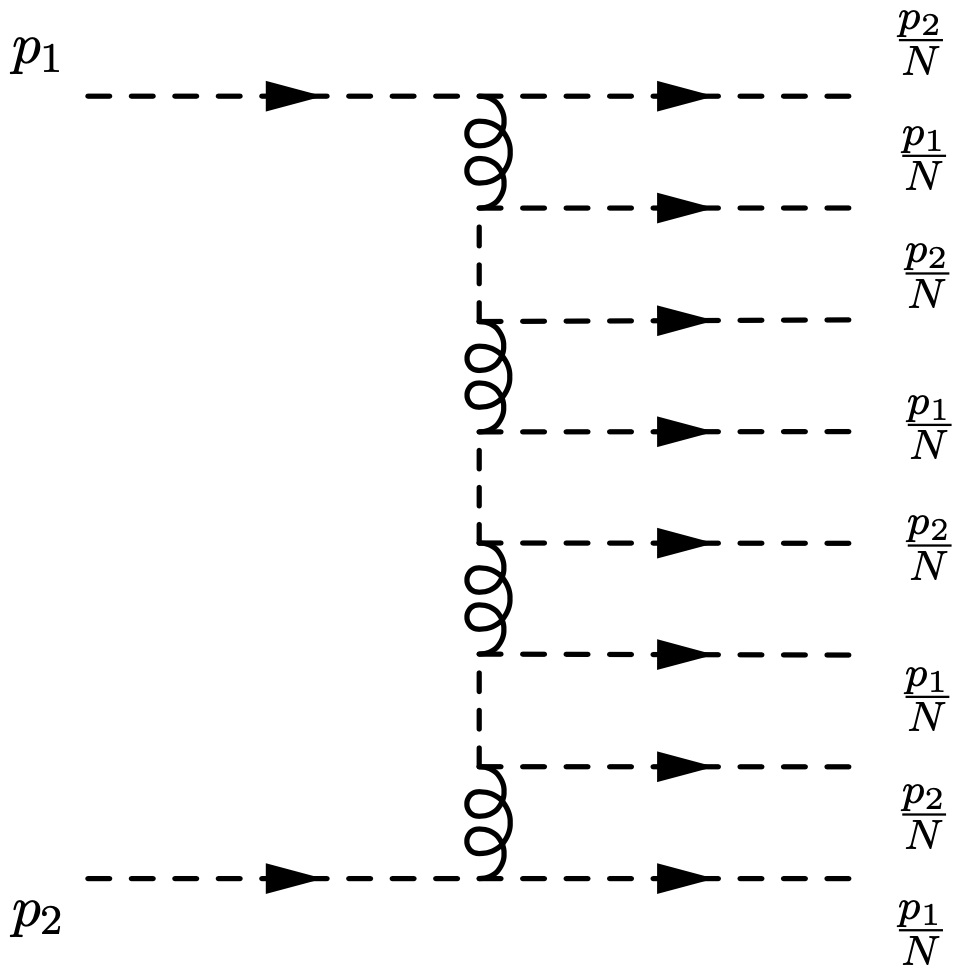}
     \caption{A $2\rightarrow 2N$ tree level diagram with democratically distributed outgoing momenta. There are two possibilities in this diagram; one where the momentum $\frac{p_{2}}{N}$ is emitted first as depicted, and the other where $\frac{p_{1}}{N}$ is first emitted. We sum both contributions. Other possible orders are forbidden by the Feynman rules and choice of incoming momenta.}
    \label{fig:2-2N_tree}
\end{figure}

Using the graviton propagator we just derived, tree level $2\rightarrow 2N$ diagrams can now be calculated. In \figref{fig:2-2N_tree}, the first outgoing leg is shown to carry momentum $\frac{p_{2}}{N}$. A similar diagram where the first emitted leg carries $\frac{p_{1}}{N}$ also exists and needs to be accounted for. 
Following the conventions of \cite{Gaddam:2020rxb, Gaddam:2020mwe}, the centre of mass energy of the collision is $s = -\frac{1}{2}\left(p_{1} + p_{2}\right)^{2}$ where we choose $p_{1} = \left(p_{1,u},0\right)$ and $p_{2} = \left(0, p_{2,v}\right)$ for the incoming particles in the Feynman diagram. Starting at the top of the diagram, and assuming the case that the first outgoing momentum is $\frac{p_{1}}{N}$, we first note that the vertices around the $n^{\text{th}}$ graviton propagator carry legs with momenta
\begin{align}
    p_i ~ &= ~ \left(\dfrac{\left(N-n+1\right)}{N}  p_{1,u} , \, \dfrac{-\left(n-1\right)}{N} p_{2,v}\right) , \,  \qquad && p_j ~ = ~ \left(\dfrac{1}{N} p_{1,u} , \, 0 \right) \, , \qquad \qquad \label{eqn:12_ij_momenta} \\
    p_l ~ &= ~ \left(\dfrac{\left(N-n\right)}{N} p_{1,u}, \, \dfrac{- n}{N} p_{2,v}\right) , \,  \qquad && p_k ~ = ~ \left(0 , \, \dfrac{1}{N} p_{2,v} \right) \, . \label{eqn:12_kl_momenta}
\end{align}
Using the graviton propagator \eqref{eqn:monopoleProp}, we then compute the quantity
\begin{align}
\mathcal{S}^{tree}_{(12)}\left(n\right) ~ \coloneqq ~ - 2 i \gamma^{2} p_{i,a} p_{j,b} \mathcal{P}^{abcd} \left(-p_{l,c}\right) p_{k,d} ~ = ~ 2 i s^{2} R^{2} \gamma^{2} \dfrac{n \left(N - n +1 \right)}{N^{4}}
\end{align}
which includes the coupling constant arising from the vertex. Similarly, for the case where the first outgoing leg carries momentum $\frac{p_{2}}{N}$, we have
\begin{align}
    p_i ~ &= ~ \left(\dfrac{\left(N-n+1\right)}{N}  p_{1,u} , \, \dfrac{-\left(n-1\right)}{N} p_{2,v}\right) , \,  \qquad && p_j ~ = ~ \left(0 , \, \dfrac{1}{N} p_{2,v} \right) \, , \qquad \qquad \label{eqn:21_ij_momenta} \\
    p_l ~ &= ~ \left(\dfrac{\left(N-n\right)}{N} p_{1,u}, \, \dfrac{- n}{N} p_{2,v}\right) , \,  \qquad && p_k ~ = ~ \left(\dfrac{1}{N} p_{1,u} , \, 0 \right) \, . \label{eqn:21_kl_momenta}
\end{align}
These lead to
\begin{align}
\mathcal{S}^{tree}_{(21)}\left(n\right) ~ \coloneqq ~ - 2 i \gamma^{2} p_{i,a} p_{j,b} \mathcal{P}^{abcd} \left(-p_{l,c}\right) p_{k,d} ~ = ~ 2 i s^{2} R^{2} \gamma^{2} \dfrac{\left(N - n\right) \left(n -1 \right)}{N^{4}} \, .
\end{align}
Therefore, we have
\begin{align}
S^{tree}\left(n\right) ~ \coloneqq ~ \mathcal{S}^{tree}_{(12)}\left(n\right) + \mathcal{S}^{tree}_{(21)}\left(n\right) \, .
\end{align}
The internal $m^{\text{th}}$ matter propagator carries momentum 
\begin{equation}
    p_m ~ = ~ \left(\dfrac{N-m}{N} p_{1,x}, \, \dfrac{- m}{N} p_{2,y}\right) \quad \implies \quad p^2_m ~ = ~ \dfrac{2 m \left(N-m\right) s}{N^2} \, .
\end{equation}
Therefore, we find
\begin{align}
    \dfrac{-i}{p^2_m + \frac{1}{R^2}} ~ &= ~ \dfrac{N^2}{2 s} \left(\dfrac{-i}{m\left(N - m\right) + \frac{N^{2}}{2 s R^{2}} }\right) ~ \approx ~ \dfrac{N^2}{2 s} \left(\dfrac{i}{m\left(N - m\right) }\right) \, ,
\end{align}
where in the last step, we observe that $m \left(N-m\right)$ always dominates $\frac{N^{2}}{2 s R^{2}}$. Piecing all the above together, the amplitude can be written as
\begin{align}\label{eqn:general_tree_amplitude}
i \mathcal{M}^{tree}_{2\rightarrow 2N} ~ &= ~ \prod_{n=1}^{N} S^{tree}\left(n\right) \prod_{m=1}^{N-1} \dfrac{N^2}{2 s} \left(\dfrac{-i}{m\left(N - m\right) }\right) \nonumber \\
&= ~ \dfrac{2is}{N! N!} \left(\dfrac{s \kappa^{2}}{N^2}\right)^N \prod\limits_{n=1}^N \biggr(n \left(N - n + 1\right) + \left(N - n\right)\left(n - 1\right)\biggr) \, .
\end{align}
Using Stirling's approximation, this amplitude can be evaluated to find
\begin{equation}\label{eqn:leading_tree_amplitude}
i \mathcal{M}^{tree}_{2\rightarrow 2N} ~ \sim ~ \dfrac{2 i s}{\pi N} \left(\dfrac{2 s \kappa^{2}}{N^2}\right)^N  \, .
\end{equation}
Owing to the democratic choice of external momenta, exchanging all the $p_{1}$ legs among themselves results in identical amplitudes. The same is true for exchanging all the $p_{2}$ legs among themselves. Summing over these diagrams gives an additional factor of $N! N!$. However, among these, there are $N!$ topologically identical ones which involve shuffling the internal graviton legs among each other. Therefore, the total tree level amplitude including the topologically distinct diagrams is
\begin{equation}\label{eqn:final_tree}
i \mathcal{M}^{tree}_{2\rightarrow 2N} ~ \sim ~ \dfrac{2 i s}{\pi N} \left(\dfrac{2 s \kappa^{2}}{N^2}\right)^N N! \quad \text{which has a maximum at} \quad N_{max} ~ \sim ~ s \kappa^{2} \, .
\end{equation}
Moreover, the amplitude \eqref{eqn:final_tree} is a sharply peaked Gaussian about the maximum value with a standard deviation of $\sqrt{\frac{N_{max}}{2\pi}}$, which validates a large $N$ approximation. As we will show in \secref{sec:timedelay}, the centre of mass energy will be related to Schwarzschild energy via $E^{2} \sim s$. Furthermore, choosing $E=M_{BH}$ is consistent with the expectation that $N_{max}$ is related to the number operator associated with the entropy of the black hole \cite{Dvali:2014ila, Ghosh:2016fvm}:
\begin{equation}
N_{max} ~ \sim ~ G E^{2} ~ \sim ~ G M^{2}_{BH} ~ \sim ~ S_{BH} \, .
\end{equation}
Additionally, we see that $N_{max} \sim E^{2} / M^{2}_{Pl}$ which indicates that for small collision energies, the elastic $2\rightarrow 2$ amplitude is the dominant one. Whereas particle production becomes important for large collision energies. Therefore, while the scattering particles may scatter with any centre of mass energies, the case of $E=M_{BH}$ may be seen as the most significant one. 

The importance of $E=M_{BH}$ was already noticed in the first quantised scattering matrix approach of 't Hooft's, which in the present formulation corresponds to $2\rightarrow 2$ scattering. When $E = M_{BH}$ in that formulation, the corresponding Shapiro shift is of the order of the Schwarzschild radius. This in turn means that the particles may or may not be shifted into the horizon because of backreaction; both possibilities need to be accounted for.\footnote{We thank Gerard 't Hooft for clarifications on this point.}

In \cite{Dvali:2014ila}, it was argued that gravity, owing to the presence of black holes, allows for a regime of classicalisation where ultra-Planckian strongly coupled dynamics is to be replaced by weakly coupled tree level processes with a large number of outgoing states. Their proposed emergent weak coupling constant is $\alpha \sim \frac{s G}{N^{2}}$ which leads to the tree level $2\rightarrow N$ amplitude \eqref{eqn:final_tree}
\begin{equation}
\sigma_{2\rightarrow N} ~ \sim ~ \alpha^{N} N! ~ = ~ \left(\dfrac{s G}{N^2}\right)^N N! \, .
\end{equation}
Moreover, it motivates the natural definition of an 't Hooft-like coupling constant
\begin{equation}
\lambda ~ \coloneqq ~ \alpha N ~ = ~ \dfrac{s G}{N} \, .
\end{equation}
The equivalence of our amplitude \eqref{eqn:final_tree} with theirs\footnote{See eq (1.3) of \cite{Dvali:2014ila} for instance.} shows that the origin of their proposed (weak) coupling constant lies in the dynamics of the near horizon region. We have explicitly derived the effective field theory with the said weakly-coupled dynamics. Our derivation naturally allows for explicit calculations of various classical and quantum corrections to the tree-level amplitude.

\subsection{The `eikonal propagator'}
\begin{figure}[h!]
\centering
   \includegraphics[scale=0.38]{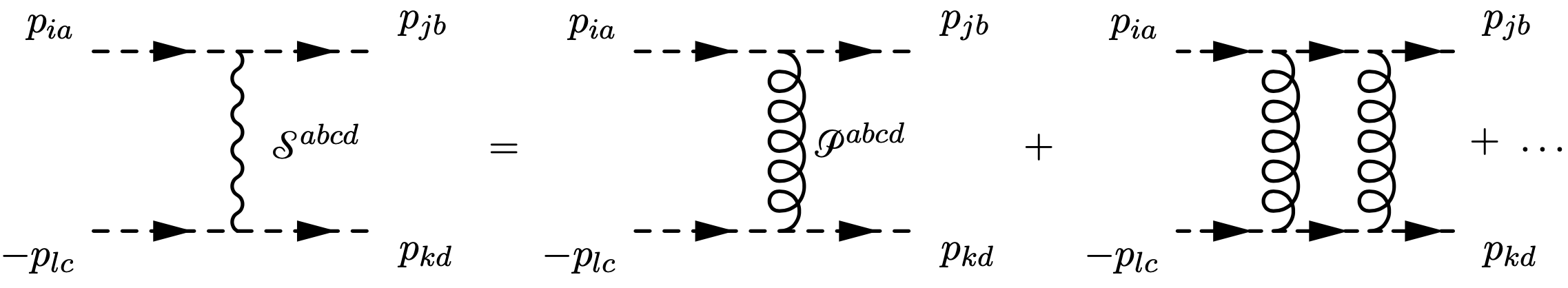}
    \caption{Each $2\rightarrow 2$ ladder can be written in the form of a modified `eikonal' propagator $\mathcal{S}^{abcd}$ that captures the resummation of all ladder exchanges entirely.}
    \label{fig:loop_summation}
\end{figure}

In this section we will calculate an infinite set of corrections to the above tree amplitude \eqref{eqn:leading_tree_amplitude}, arising from all possible graviton exchanges as shown in \figref{fig:2-2N_tree}. The main idea to calculate the ladder of ladders diagrams involves rewriting every $2\rightarrow 2$ ladder in the diagram in terms of an eikonal graviton propagator $\mathcal{S}$ defined via
\begin{align}
    \mathcal{S}_{ijlk}(q) ~ &\coloneqq ~ - 2 i \gamma^2 \int \dfrac{\ed^2 k}{\left(2 \pi \right)^2} p_{ia} p_{jb} \mathcal{P}^{abcd} \left(k\right) \left(- p_{lc}\right) p_{kd}\int \ed^2 x e^{i (k-q)\cdot x} \dfrac{e^{i\chi_{ijlk}} - 1}{\chi_{ijkl}} \, , \label{eqn:generalS} \\
    \chi_{ijkl} ~ &\coloneqq ~ - i \gamma^2\int\frac{\ed^2 k}{(2\pi)^2} p_{ia} p_{jb} \left(2 i \mathcal{P}^{abcd}(k)\right) \left(-p_{lc}\right) p_{kd} e^{- i k \cdot x} \nonumber \\
    &\qquad \qquad \qquad \qquad \qquad \qquad \qquad \times d \left(k, p_j, p_i\right) d\left(- k, p_k, - p_l\right) \, , \label{eqn:generalChi} \\
    d(k,p',p) ~ &\coloneqq ~ \frac{1}{2 p'\cdot k+i\epsilon}-\frac{1}{2p\cdot k-i\epsilon} \, . \label{eqn:d}
\end{align}
following the standard assumptions of \cite{Levy:1969cr, Kabat:1992tb}. At tree level, $\chi = 0$ and the propagator $\mathcal{S}$ reduces to the tree level propagator that results in \figref{fig:2-2N_tree}. Pictorially, one may think of this propagator $\mathcal{S}$ as shown in \figref{fig:loop_summation}; so, $\mathcal{S}$ is a propagator that captures the resummation of the eikonal $2\rightarrow 2$ graphs.
The resulting amplitudes arising from this eikonal propagator, therefore, need the integration of \eqref{eqn:generalChi}, which is a sum of four terms. Each of these terms is of the form
\begin{equation}\label{eqn:I}
I ~ = ~ \int \dfrac{\ed^2 k}{(2\pi)^2} e^{- i k_{u} u - i k_{v} v} f\left(k^{2}\right) \left(\dfrac{1}{2 p_{i} \cdot k + i \sigma_{i} \epsilon} ~ \times ~ \dfrac{1}{2 p_{j} \cdot k + i \sigma_{j} \epsilon}\right) \, ,
\end{equation}
where $\sigma_{i,j} = \pm$ and the function $f\left(k^{2}\right)$ is assumed to be regular. To solve this integral, we first write $p_{i} = \left(A_{i} p_{i,u} , B_{i} p_{i,v}\right)$ and $p_{j} = \left(A_{j} p_{j,u} , B_{j} p_{j,v}\right)$. Then we define 
\begin{equation}
    w ~ \coloneqq \begin{pmatrix} w_u \\ w_v \end{pmatrix}
    ~ \coloneqq ~ \mathbb{M}_{(ij)} \begin{pmatrix} k_u \\ k_v \end{pmatrix} 
    \coloneqq  \begin{pmatrix} B_i p_{j,v} & A_i p_{i,u} \\ B_j p_{j,v} & A_j p_{i,u} \end{pmatrix} \begin{pmatrix} k_u \\ k_v \end{pmatrix} ~ \text{and} ~ \begin{pmatrix} q_u \\ q_v \end{pmatrix} \coloneqq \left(\mathbb{M}^{-1}_{(ij)}\right)^{T} \begin{pmatrix} u \\ v \end{pmatrix}
\end{equation}
with $k \cdot x = w \cdot q$. The label $(ij)$ on the matrix $\mathbb{M}$ keeps track of which of the four terms in \eqref{eqn:generalChi} is under consideration. This allows us to reduce the integral \eqref{eqn:I} to
\begin{equation}
I ~ = ~ \dfrac{1}{4 \left|\det\mathbb{M}_{(ij)}\right|} \int \dfrac{\ed^2 w}{(2\pi)^2} e^{- i w_{u} q_{u} - i w_{v} q_{v}} f\left(\left(\mathbb{M}^{-1}_{(ij)} w\right)^{2}\right) \left(\dfrac{1}{w_{u} - i \sigma_{i} \epsilon} ~ \times ~ \dfrac{1}{w_{v} - i \sigma_{j} \epsilon}\right) \, .
\end{equation}
where the double poles in \eqref{eqn:I} are reduced to single poles. A straightforward application of the residue theorem then results in
\begin{align}
I ~ &= ~ \dfrac{\sigma_{i} \left(2\pi i\right)}{4 \left|\det \mathbb{M}_{(ij)}\right|} \dfrac{\Theta_{(ij)}\left(-\sigma_{i} q_{u}\right)}{2\pi} \int \dfrac{\mathrm{d}w_{v}}{2\pi} e^{-iw_{v}q_{v}} f\left(\left(\mathbb{M}^{-1}_{(ij)} w|_{w_{u}=0}\right)\right) \dfrac{1}{w_{v} - i \sigma_{j} \epsilon} \nonumber \\
&= ~ \dfrac{\sigma_{i} \sigma_{j} \left(2 \pi i\right)^{2}}{4 \left|\det \mathbb{M}_{(ij)}\right|} \dfrac{f\left(0\right)}{\left(2\pi\right)^{2}} \Theta_{(ij)}\left(- \sigma_{i} q_{u}\right) \Theta_{(ij)}\left(-\sigma_{j} q_{v}\right) \, ,
\end{align}
where $\Theta$ is the Heaviside step function (which inherits the label $(ij)$ from its argument $q$ which contains the label implicitly via the matrix $\mathbb{M}$) which we choose to be symmetric with $\Theta\left(0\right) = \frac{1}{2}$; it is the unique choice of convention that agrees with the correct $2\rightarrow 2$ eikonal amplitude. Using this method, the eikonal function in \eqref{eqn:generalChi} can be written as
\begin{align}
\chi_{ijkl}\left(x\right) ~ &= ~ C_{ijkl} \dfrac{s}{N^{2}} \left(\dfrac{\Theta_{(jk)}\left(- q_{u}\right) \Theta_{(jk)}\left(q_{v}\right)}{\left|\det \mathbb{M}_{(jk)}\right|} + \dfrac{\Theta_{(jl)}\left(- q_{u}\right) \Theta_{(jl)}\left(q_{v}\right)}{\left|\det \mathbb{M}_{(jl)}\right|} + \dfrac{\Theta_{(ik)}\left(q_{u}\right) \Theta_{(ik)}\left(q_{v}\right)}{\left|\det \mathbb{M}_{(ik)}\right|} \right. \nonumber \\
&\qquad \qquad \qquad \qquad \left. \dfrac{\Theta_{(il)}\left(q_{u}\right) \Theta_{(il)}\left(q_{v}\right)}{\left|\det \mathbb{M}_{(il)}\right|}\right) \, ,
\end{align}
where we defined the function
\begin{equation}
C_{ijkl} ~ = ~ \dfrac{\gamma^{2} N^{2}}{2 s} p_{ia} p_{jb} \mathcal{P}^{abcd} \left(k\right) \left(- p_{lc}\right) p_{kd} \, .
\end{equation}
This method employed so far is valid for any spherically symmetric background. However, it turns out to be useful only in the present case because of the momentum independence of the monopole mode propagator in \eqref{eqn:monopoleProp}. This allows us to integrate \eqref{eqn:generalS}, by first performing the $k$ integral which results in a $\delta^{(2)}\left(x\right)$, simplifying the $x$ integral. We find
\begin{equation}
\mathcal{S}_{ijkl} ~ = ~ \dfrac{16 s}{N^{2}} C_{ijkl} \dfrac{e^{i \chi_{ijkl}\left(0\right)} - 1}{\chi_{ijkl}\left(0\right)} \, .
\end{equation}
Just as we did in the tree-level calculation in \secref{sec:tree_saddle}, we will treat the external particles to be on-shell even as they are chosen to be light-like. In this case, it may be checked that $C_{ijkl}=0$ if $p_{j} = p_{k}$. We will now evaluate the eikonal propagator $\mathcal{S} \coloneqq \mathcal{S}^{(12)} + \mathcal{S}^{(21)}$ which correspond to the two cases depending on the order of the emitted outgoing legs in the Feynman diagram. Starting at the top of the diagram in \figref{fig:2-2N_ladders}, we will follow the same strategy as in \secref{sec:tree_saddle}. Using the momenta for the legs around each ladder as in \eqref{eqn:12_ij_momenta} and \eqref{eqn:12_kl_momenta}, we find
\begin{equation}\label{eqn:chip1p2}
\chi^{(12)}\left(n\right)~ \coloneqq ~ C^{(12)}\left(n\right) F^{(12)}\left(n\right) \, ,
\end{equation}
where
\begin{equation}
C^{(12)}\left(n\right) ~ = ~ \dfrac{s \kappa^{2}}{8} \dfrac{n \left(N - n + 1\right)}{N^{2}} \quad \text{and} \quad F^{(12)}\left(n\right) ~ = ~ \left(1 + \dfrac{1}{n} + \dfrac{1}{N - n + 1} + \dfrac{1}{N}\right) \, .
\end{equation}
Once the momenta are explicitly specified, the labels $ijkl$ are no longer necessary and have therefore been promptly dropped. Similarly, using \eqref{eqn:21_ij_momenta} and \eqref{eqn:21_kl_momenta} we find 
\begin{equation}\label{eqn:chip2p1}
\chi^{(21)}\left(n\right) ~ \coloneqq ~ C^{(21)}\left(n\right) F^{(21)}\left(n\right) \, ,
\end{equation}
where
\begin{equation}
C^{(21)}\left(n\right) ~ = ~ \dfrac{s \kappa^{2}}{8} \dfrac{\left(N - n\right) \left(n - 1\right)}{N^{2}} \quad \text{and} \quad F^{(21)}\left(n\right) ~ = ~ \left(1 + \dfrac{1}{n - 1} + \dfrac{1}{N - n} + \dfrac{1}{N}\right) \, .
\end{equation}
Combining \eqref{eqn:chip1p2} and \eqref{eqn:chip2p1}, we find the final eikonal graviton propagator
\begin{equation}
\mathcal{S}\left(n\right) ~ = ~ \dfrac{16 s}{N^{2}} \left(\dfrac{e^{i \chi^{(12)}\left(n\right)} - 1}{F^{(12)}\left(n\right)} + \dfrac{e^{i \chi^{(21)}\left(n\right)} - 1}{F^{(21)}\left(n\right)}\right) \, .
\end{equation}
Finally, the scalar momenta and propagators are clearly unchanged from the tree level case:
\begin{align}
    \prod_{m=1}^{N-1} \dfrac{-i}{p^2_m + \frac{1}{R^2}} ~ \approx ~ \prod_{m=1}^{N-1} \dfrac{N^{2}}{2 s} \dfrac{- i}{m\left(N - m\right) } ~ = ~ \left(\dfrac{- i N^{2}}{2 s}\right)^{N-1} \dfrac{1}{\left(N - 1\right)! \left(N - 1\right)!} \, .
\end{align}

\subsection{The eikonal $\mathcal{M}_{2\rightarrow 2N}$ amplitude}
The all-loop `ladder of ladders' amplitude evaluates to
\begin{align}\label{eqn:ladder_of_ladders}
i\mathcal{M}_{2\rightarrow 2N} ~ &= ~ \prod_{n=1}^{N} \mathcal{S}\left(n\right) \prod_{m=1}^{N-1} \dfrac{-i}{p^{2}_{m} + R^{-2}} \nonumber \\
&= ~ \left(- 8 i\right)^{N} \dfrac{4 i s}{N!} \prod_{n=1}^{N} \left(\dfrac{e^{i \chi^{(12)}\left(n\right)} - 1}{F^{(12)}\left(n\right)} + \dfrac{e^{i \chi^{(21)}\left(n\right)} - 1}{F^{(21)}\left(n\right)}\right) \, ,
\end{align}
where we have again incorporated a factor of $2 \left(N!\right)$ to account for all the topologically distinct diagrams as in the tree level amplitude. While evaluating the above product exactly is difficult, there are two quick consistency checks one may immediately look for. First, when $n=1=N$, we note that $F^{(12)}\rightarrow 4$ and $F^{(21)}\rightarrow \infty$ and $\chi^{(12)}=\frac{s \kappa^{2}}{2}$, resulting in the $2\rightarrow 2$ amplitude
\begin{equation}\label{eqn:2-2eikonal}
i \mathcal{M}_{2\rightarrow 2} ~ = ~ 4 s \bigr(\exp\left(4 i \pi G_{N} s\right) - 1\bigr) \, ,
\end{equation}
which agrees with the amplitude found in \cite{Gaddam:2020rxb, Gaddam:2020mwe} with $\ell = 0$ and $\hbar = 1$. Second, keeping only the leading order part of the oscillatory exponentials in the eikonal functions to write $e^{i \chi} \approx 1 + i \chi$, we immediately see that the amplitude \eqref{eqn:ladder_of_ladders} matches the tree level amplitude in \eqref{eqn:general_tree_amplitude}.

\section{Eisenbud-Wigner time delay}\label{sec:timedelay}
In order to understand a notion of time delay, we first need to specify the coordinates and corresponding energies with respect to which it is measured. We call the energy operator in global Schwarzschild coordinates as $E = - i \partial_{t} = - \frac{i}{R} \partial_{\tau}$. In Kruskal coordinates, this can be written as $E = - \frac{i}{R} \left(U \partial_{U} - V \partial_{V}\right) = \frac{1}{R} \epsilon^{ab} x_{a} p_{a}$, where we defined $- i \partial_{a} = p_{a}$. For the individual infalling particles, this implies $E^{2} = E_{1} E_{2} = - \frac{1}{R^{2}} U_{1} V_{2} p_{(1,U)} p_{(2,V)} = \frac{s}{R^{2}} U_{1} V_{2}$. Using the tortoise coordinate $r^{\star} = r + R \log\left|\frac{r}{R} - 1\right|$, we first observe that
\begin{equation}
U ~ = ~ \sqrt{\dfrac{2 R^{2}}{e}} \exp\left(\dfrac{r^{\star} + t}{2 R}\right) \quad \text{and} \quad V ~ = ~ \sqrt{\dfrac{2 R^{2}}{e}} \exp\left(\dfrac{r^{\star} - t}{2 R}\right) \, .
\end{equation}
The infalling particles, by definition of the scattering problem, originate from the asymptotic past defined by $u = \frac{r^{\star} + t}{2R}$ and $v = \frac{r^{\star} - t}{2R}$, with $r^{\star}_{1} \rightarrow -\infty$, $r^{\star}_{2} \rightarrow \infty$, and $t_{1,2} \rightarrow - \infty$ as can be seen from \figref{fig:2-2N_schematic}. Piecing these together, we have $E^{2} = \frac{2 s}{e}$. Given Wigner's time delay defined in \eqref{eqn:wigner_timedelay} in Kruskal coordinates, the corresponding time delay in Schwarschild coordinates can be inferred from the above coordinate transformations to find
\begin{align}\label{eqn:wigner_schwarzschild}
\Delta t ~ &= ~ 2 R \cosh^{-1}\left(\dfrac{1}{\sqrt{2}} \text{Re}\left[- i S^\dagger \dfrac{\partial S}{\partial E}\right]\right) \, .
\end{align}

\subsection{Elastic $2\rightarrow 2$ scattering and scrambling time}
Using the relation between $E$ and $s$ as above, and applying the time delay expression \eqref{eqn:wigner_schwarzschild} to the $2\rightarrow 2$ black hole eikonal amplitude \eqref{eqn:2-2eikonal}, we find
\begin{align}
\Delta t ~ &\sim ~ G M_{BH} \log \left(G E\right) + \dots \nonumber \\
&= ~ G M_{BH} \log \left(G M_{BH}\right) + \dots  \, ,
\end{align}
where we set $E = M_{BH}$ as explained at the end of \secref{sec:tree_saddle} and approximated $\cosh^{-1}(x) \approx \log(2x)$. This is the familiar scrambling time indicating that the $2\rightarrow 2$ eikonal amplitude probes elastic near horizon properties of the horizon \cite{Shenker:2013pqa, Shenker:2014cwa, Maldacena:2015waa}. However, it does not probe physics associated with Page time which is crucial for information retrieval. Notice that $E=M_{BH}$ is necessary for consistency with these references. We will make further comments on this in \secref{sec:outlook}. 

\subsection{Particle production and Page time}
The multi-channel generalisation of the Eisenbud-Wigner time delay \eqref{eqn:wigner_schwarzschild} for inelastic scattering is given by \cite{Martin:1981dea}
\begin{align}\label{eqn:wigner_schwarzschild_multichannel}
\Delta t ~ &= ~ 2 R \cosh^{-1}\left(\dfrac{1}{\sqrt{2}} \text{Re}\left[- i \sum_{N=1}^{\infty} S^\dagger_{2\rightarrow 2N} \dfrac{\partial S_{2\rightarrow 2N}}{\partial E}\right]\right) \, .
\end{align}
This may be understood as the sum over times that the scattering particles spend in each `region of configuration space'. To evaluate this, we first use Stirling's approximation on \eqref{eqn:final_tree} to find
\begin{equation}
i \mathcal{M}^{tree}_{2\rightarrow 2N} ~ \sim ~ i s \left(\dfrac{2 s \kappa^{2}}{e^2}\right)^N \dfrac{2 \pi N}{N!} \, .
\end{equation}
Writing the scattering matrix associated with this amplitude as $S_{2\rightarrow 2N} = 1 + i \mathcal{M}_{2\rightarrow 2N}$, we see that 
\begin{align}
\text{Re}\left[- i \sum_{N=1}^{\infty} S^\dagger_{2\rightarrow 2N} \dfrac{\partial S_{2\rightarrow 2N}}{\partial E}\right] ~ \sim ~ \exp\left(E^{2} \kappa^{2}\right) \, ,
\end{align}
where we used that $E^{2} \sim s$ is very large. Therefore, the time delay is now given by
\begin{align}\label{eqn:pagetime}
\Delta t ~ &\approx ~ G M_{BH} \, E^{2} \, \kappa^{2} \nonumber \\
&\approx ~ G^{2} M^{3}_{BH} \, ,
\end{align}
where we approximated the $\cosh^{-1}$ in \eqref{eqn:wigner_schwarzschild_multichannel} to a logarithm for large energies, and in the second line, we have again used $E = M_{BH}$ as explained at the end of \secref{sec:tree_saddle}. The single channel time delay at the maximum value of $N$ also yields the Page time. Of course, the oscillations in the eikonal amplitude \eqref{eqn:ladder_of_ladders} will produce additional contributions to this time delay, but these will be logarithmic in the energy and therefore subleading. Whereas the dominant contribution arises from the exponential growth of the tree level probabilities.

\section{Outlook}\label{sec:outlook}
In this article, we have explicitly calculated an infinite number of $2\rightarrow 2N$ `ladder of ladders' diagrams mediated by virtual gravitons, to all orders in the coupling constant $M_{Pl}/M_{BH}$. While there are several other diagrams that may possibly contribute to $2\rightarrow 2N$ scattering, the aim of the article was to provide an explicit calculation of detailed black hole scattering matrix elements that capture particle production. This is important for considerations of the information paradox, as is reflected by the time delay \eqref{eqn:pagetime} that measures how long one has to wait before information retrieval is possible. It is remarkable that the lifetime associated with the intermediate state of the black hole emerges from the scattering matrix. 

Page time may be thought of as the time it takes for a black hole (formed by collapse in a Minkowski background) to lose half its mass. The reason for this time scale to appear in the study of scattering on a black hole background may be understood as follows. Lacking even the classical time dependent solution describing the collapse, we posed and answered the following question in the present article: if we threw two particles (with centre of mass energy $E=M_{BH}$) into an existing black hole, thereby doubling its mass, how long would it take for the black hole to then lose half its mass (equivalently, how long would it take for all the ingoing energy to exit the black hole)? Remarkably, the answer turned out to be Page time. It need not have. This lends credibility to our claim that our approach indeed reveals the dynamics that allows for information retrieval from Hawking radiation. Phrased this way, it is clear that $E=M_{BH}$ is special. For any other $E$, momentum conservation in the Feynman diagrams would not correspond to the black hole losing half its mass. 

Alternatively, one may consider the collapsing problem in flat space instead of scattering in a black hole background. It is well known that an apparent horizon forms well before the infalling matter crosses the eventual Schwarzschild radius. Therefore, in the final stages of collapse, matter would essentially be falling into an apparent horizon that is very closely approximated by the event horizon. While this approximation is very poor in the initial stages of collapse, it is a very good one towards the final stages of collapse. In fact, this may be thought of as the reason why Hawking's analysis was completely based on a fixed black hole background, even though the claims concern a collapsing and evaporating black hole.

Our results leave a lot of questions to be explored, several of which are within technical reach in the black hole eikonal phase of quantum gravity where scattering energies satisfy $E M_{BH} \gg M^{2}_{Pl}$. For instance, sub-leading corrections of various kinds can be calculated order by order in perturbation theory, possibly providing a window into the ultraviolet, in the spirit of \cite{Amati:1987wq, Amati:1987uf, Amati:1992zb}. In the same vein, the black hole/string transition is an interesting avenue to explore \cite{Veneziano:2012yj}.

\section*{Acknowledgements}
It is a pleasure to thank Panos Betzios, Bernard de Wit, Gia Dvali, Umut Gursoy, Gerard 't Hooft, Anna Karlsson, Olga Papadoulaki, and Gabriele Veneziano for various helpful conversations. This work is supported by the Delta-Institute for Theoretical Physics (D-ITP) that is funded by the Dutch Ministry of Education, Culture and Science (OCW).

\printbibliography
\end{document}